\UseRawInputEncoding
\documentclass[preprintnumbers,superscriptaddress,amsmath,amssymb,prd,preprint,nofootinbib]{revtex4-2}
\usepackage{amssymb}
\usepackage{mathrsfs}
\usepackage{amsthm}
\usepackage{graphicx}
\usepackage{amsfonts}
\usepackage[colorlinks,linkcolor=red,anchorcolor=blue,citecolor=green]{hyperref}
\usepackage{subfigure}
\usepackage{float}
\usepackage{enumerate}

\begin{document}	
\title{Rotating black holes in de Rham-Gabadadze-Tolley massive gravity: Newman-Janis Algorithm}
\author{Ping Li}
\email[]{Lip57120@huas.edu.cn}
\affiliation{College of Mathematics and Physics $\&$ Hunan Province Key Laboratory Integration and Optical Manufacturing Technology, Hunan University of Arts and Sciences, 3150 Dongting Dadao, Changde City, Hunan Province 415000, China}
\author{Jiang-he Yang}
\email[]{yjianghe@163.com}
\affiliation{College of Mathematics and Physics $\&$ Hunan Province Key Laboratory Integration and Optical Manufacturing Technology, Hunan University of Arts and Sciences, 3150 Dongting Dadao, Changde City, Hunan Province 415000, China}
\affiliation{Center for Astrophysics, Guangzhou University, 230 West Ring Road, Guangzhou, Guangdong Province 510006, China}

\begin{abstract}
We report the discovery of rotating black hole solutions within the framework of de Rham-Gabadadze-Tolley (dRGT) massive gravity. We demonstrate that any nonunitary gauge with the Minkowski reference metric are equal to a unitary gauge with some curved reference metric. Based on this Lemma, we revisit the process of deriving black hole solutions in dRGT theory. We explain how to obtain a static, spherically symmetric solution and then transform it into the corresponding rotating black hole using the Newman-Janis algorithm. For the first time, we provide an analytic expression for a hairy black hole that can reduce to the non-rotating case. Additionally, we confirm that the Newman-Janis algorithm is applicable in the context of massive gravity.
\end{abstract}
\maketitle
	
\emph{Introduction.} In this letter, we aim to address three questions.

1. What is the rotating black hole solution that can reproduce the static spherical symmetry (SSS) case in massive gravity? Traditionally, Einstein's gravity has been recognized as the only massless spin-2 theory. For the massive extension, the dRGT theory is the most competitive candidate \cite{deRham2011}. In this theory, the primary contribution to the decoupling limit is a total derivative, which prevents the emergence of ghosts. Beyond the decoupling limit, the lapse function acts as a Lagrange multiplier, constraining the sixth mode in an arbitrary background. Consequently, the dRGT theory describes a massive graviton that propagates 5 healthy degrees of freedom. A comprehensive review of massive gravity can be found in the literature \cite{deRham2014}. Numerous papers have explored black hole solutions within the dRGT framework \cite{Nieuwenhuizen2011,Cai2013,Berezhiani2011,Li2016-1,Cai2015,Jafari2017}. However, these solutions have primarily been limited to static spherical symmetry, making it challenging to extend to rotating cases. The introduction of nonlinear terms in Lagrangian significantly complicates the calculation of black hole solutions, especially for rotating scenarios. We first present an analytic expression for a rotating black hole that can reproduce the SSS solution found in non-rotating cases.

2. Is the no-hair conjecture true for massive gravity theory? In Einstein's theory of gravity, the no-hair conjecture, proposed by John Wheeler \cite{Wheeler}, suggests that all black holes in nature belong to the Kerr-Newman family. For many years, researchers have sought to prove or disprove this conjecture. There are two fundamentally different approaches to testing it. First, we can examine whether the no-hair conjecture holds for Einstein gravity when coupled with standard matter fields. In the Einstein-Yang-Mills theory, Volkov and Galtsov discovered black hole solutions that exhibit Yang-Mills hair \cite{Volkov1989,Volkov1990}. In 2014, Herdeiro and Radu presented solutions for rotating black holes that possess scalar hair in the context of Einstein gravity minimally coupled with a complex, massive scalar field \cite{Herdeiro2014}. For a comprehensive review of hairy black holes, we recommend consulting the relevant literature \cite{Herdeiro2015}. Second, we explore whether the no-hair conjecture is valid in the framework of standard gravity theory. Some studies have claimed to derive hairy black holes within the dRGT theory \cite{Li2016-1,Herdeiro2015}; however, these claims often face challenges in generalizing to rotating cases or in providing analytic expressions. In light of this issue, we present an analytical expression for the hairy black hole in massive gravity.

3. Dose the Newman-Janis algorithm (NJA) is also valid for the hairy black holes in massive gravity? The NJA \cite{Newman1965} is a powerful technique used to generate Kerr or Kerr-like metrics from the corresponding spherically symmetric solutions (SSS). By employing a set of complex coordinate transformations, the NJA serves as a fast and efficient shortcut for deriving rotating solutions. With some modifications, the applicability of the NJA has become more universal. Azreg-A\"{\i}nou \cite{Azreg2014-1,Azreg2014-2} extended the NJA to derive regular rotating black holes, while Chaturvedi \textit{et. al.} \cite{Chaturvedi2023} adapted it for use in Einstein gravity with a cosmological constant. However, the physical mechanism behind the NJA remains poorly understood. It has been demonstrated in previous research that the NJA is not applicable in quadratic gravity models \cite{Cadoni:2011,Hansen:2013,Ayzenberg:2015}. In this letter, we show that the NJA is still valid in the dRGT massive gravity framework. It is important to note that the NJA operates directly on the solutions and does not depend on the Einstein equations. Therefore, all solutions obtained through this method must be verified to ensure they satisfy the relevant field equations. Additionally, there is another paper that employs a rigorous analytic calculation method to derive the same solutions.

\emph{The field equation.} We consider the dRGT-Maxwell theory with the action
\begin{equation}\label{action}
  S=\int d^4x\sqrt{-g}\left( R+m^2U(g,\phi^a)-\frac{1}{16\pi}F_{\alpha\beta}F^{\alpha\beta} \right),
\end{equation}
where $R$ is the Ricci scalar, $m$ is the graviton mass and $F_{\alpha\beta}=\partial_{\alpha}A_{\beta}-\partial_{\beta}A_{\alpha}$ is electromagnetic field tensor. The potential $U$ in four-dimensional spacetime is composed of three parts,
\begin{equation}\label{potential}
 U(g,\phi^a)=U_2+\alpha_3U_3+\alpha_4U_4,
\end{equation}
where $\alpha_3,\alpha_4$ are dimensionless parameters and
\begin{align}
  U_2&= [\mathcal{K}]^2-[\mathcal{K}^{2}],  \\
  U_3& =[\mathcal{K}]^3-3[\mathcal{K}][\mathcal{K}^{2}]+2[\mathcal{K}^{3}],\\
  U_4&=[\mathcal{K}]^4-6[\mathcal{K}]^2[\mathcal{K}^{2}]+8[\mathcal{K}][\mathcal{K}^{3}]+3[\mathcal{K}^{2}]^2-6[\mathcal{K}^{4}],\\
  \mathcal{K}^{\alpha}{}_{\beta}&=\delta^{\alpha}_{\beta}-\gamma{}^{\alpha}{}_ {\beta},
\end{align}
and $\gamma^{2}{}^{\alpha}{}_ {\beta}=g^{\alpha\sigma}\partial_{\sigma}\phi^a\partial_{\beta}\phi^b\eta_{ab}$. We have chosen $G=1$.

The field equation is obtained by variating the action (\ref{action}) with respect to the metric $g_{\alpha\beta}$
\begin{equation}\label{ModEin}
 G_{\alpha\beta}+m^2T^{(\mathcal{K})}_{\alpha\beta}=8\pi T^{(m)}_{\alpha\beta},
\end{equation}
where $G_{\alpha\beta}$ is the Einstein tensor and 
\begin{align}
T^{(m)}{}^{\alpha}{}_{\beta}&=\frac{1}{4\pi}\bigg(F^{\alpha}{}_{\sigma}F_{\beta}{}^{\sigma}-\frac{1}{4}\delta^{\alpha}_{\beta}F_{\sigma \kappa}F^{\sigma\kappa}\bigg),\\
 T^{(\mathcal{K})}{}^{\alpha}{}_{\beta}&=(-\frac{U}{2}+3+6(\alpha_3+\alpha_4))\delta^{\alpha}_{\beta}+([\gamma]-3)\gamma^{\alpha}{}_{\beta}-\gamma^{2}{}^{\alpha}{}_{\beta}
-\frac{3\alpha_3}{2}(
 (6-4[\gamma]+\mathcal{U}_2)\gamma^{\alpha}{}_{\beta}\nonumber\\
 & -2([\gamma]-2)\gamma^{2}{}^{\alpha}{}_{\beta} +2\gamma^{3}{}^{\alpha}{}_{\beta})-2\alpha_4((6-6[\gamma]+3\mathcal{U}_2-\mathcal{U}_3)\gamma^{\alpha}{}_{\beta}+3(2-2 [\gamma]+\mathcal{U}_2)\gamma^{2}{}^{\alpha}{}_{\beta}\nonumber\\
 &+6(1-[\gamma])\gamma^{3}{}^{\alpha}{}_{\beta}+6\gamma^{4}{}^{\alpha}{}_{\beta} ),\label{TK}
\end{align}
with $\mathcal{U}_2 = [\gamma]^2-[\gamma^{2}],\mathcal{U}_3  =[\gamma]^3-3[\gamma][\gamma^{2}] +2[\gamma^{3}], \mathcal{U}_4=[\gamma]^4-6[\gamma^{2}][\gamma]^2+8[\gamma][\gamma^{3}] +3[\gamma^{2}]^2-6[\gamma^{4}]$. We have using the relationship $\delta [\gamma^{i}]=\frac{i}{2}\gamma^{i}_{\alpha\beta}\delta g^{\alpha\beta}$.

\emph{SSS ans\"{a}tz.} We first derive the SSS solution in dRGT theory. The SSS metric is chosen by 
\begin{equation}\label{StaticM}
  ds^2=f(r)dt^2-\frac{1}{f(r)}dr^2-r^2(d\theta^2+\sin^2\theta d\varphi^2).
\end{equation}
The non-zero Einstein tensors components are obtained by directly calculation
\begin{align}
G^0{}_0&=G^{1}{}_{1}=-\frac{f'}{r}-\frac{f-1}{r^2} ,\\
G^2{}_2&=G^3{}_3=-\frac{f''}{2}-\frac{f'}{r}.
\end{align}
For the spherically symmetric electromagnetic field, the energy-momentum tensor is well-known
\begin{align}
T^{(m)}{}^0{}_0&=T^{(m)}{}^{1}{}_{1}=\frac{Q^2}{8\pi r^4} ,\\
T^{(m)}{}^2{}_2&=T^{(m)}{}^3{}_3=-\frac{Q^2}{8\pi r^4}.
\end{align}

For the St\"{u}ckelberg fields $\phi^a$, the black hole solution with the unitary gauge $\phi^a=x^{\alpha}\delta^{a}_{\alpha}$ would encounter a problem for fluctuations \cite{Berezhiani2011}. Thus, the self-consistent black hole solution only can be derived by the nonunitary gauge. Notice that in the nonunitary gauge, there is $\gamma^{2}{}^{\alpha}{}_ {\beta}=g^{\alpha\sigma}\partial_{\sigma}\phi^a\partial_{\beta}\phi^b\eta_{ab}$. At the same time, we can release the reference metric $\eta_{ab}$ to a more general form $f_{ab}$. If choosing the unitary gauge $\phi^a=x^{\alpha}\delta^{a}_{\alpha}$ with the general reference metric $\bar{f}_{ab}$, one have $\bar{\gamma}^{2}{}^{\alpha}{}_ {\beta}=g^{\alpha\sigma}\delta_{\sigma}^a\delta_{\beta}^b\bar{f}_{ab}$. When $\bar{f}_{\alpha\beta}=\partial_{\alpha}\phi^a\partial_{\beta}\phi^b\eta_{ab}$, the two matrix $\gamma^{2}$ derived by two different methods are equal $\gamma^{2}{}^{\alpha}{}_ {\beta}=\bar{\gamma}^{2}{}^{\alpha}{}_ {\beta}$. We then have the Lemma: \emph{Arbitrary nonunitary gauge $\phi^a$ with the Minkowski reference metric $\eta_{ab}$ would always led to the unitary gauge $\bar{\phi}^a=x^{\alpha}\delta_{\alpha}^a$ with some curved reference metric $\bar{f}_{ab}$.} Instead of choosing the nonunitary gauge, we use the unitary gauge $\phi^a=x^{\alpha}\delta^{a}_{\alpha}$ with a general reference metric
\begin{equation}
  f_{ab}=\left(
           \begin{array}{cccc}
             f_{00} & f_{01} & 0 & 0 \\
             f_{01} & -f_{11} & 0 & 0 \\
             0 & 0 & -f_{22} & 0 \\
             0 & 0 & 0 & -f_{33} \\
           \end{array}
         \right).
\end{equation}
The matrix $\gamma^2{}^{\alpha}{}_{\beta}$ can be written as $\gamma^2=\left(
                                                                         \begin{array}{cc}
                                                                        \Xi_1 & 0 \\
                                                                         0 & \Xi_2 \\
                                                                         \end{array}
                                                                       \right)
$ where the submatrixs $\Xi_1$ and $\Xi_2$ have the form
\begin{equation}
 \Xi_1=\left(
         \begin{array}{cc}
           f^{-1}f_{00} &  f^{-1}f_{01} \\
           ff_{01} & ff_{11} \\
         \end{array}
       \right),\quad
  \Xi_2=\left(
         \begin{array}{cc}
           \frac{f_{22}}{r^2} & 0 \\
           0 & \frac{f_{33}}{r^2\sin^2\theta} \\
         \end{array}
       \right).
\end{equation}
Using the Cayley-Hamilton theorem, one can calculate the square root of matrix $\sqrt{\Xi_1}$ and $\sqrt{\Xi_2}$. Furthermore, we can derive the expressions for $T^{(\mathcal{K})}{}^{\alpha}{}_{\beta}$, which is shown in appendix. To make the expression more obvious, consider the following form
\begin{align}
T^{(\mathcal{K})}{}^0{}_0&=T^{(\mathcal{K})}{}^{1}{}_{1}=\frac{S^2}{ r^4}-\Lambda ,\\
T^{(\mathcal{K})}{}^2{}_2&=T^{(\mathcal{K})}{}^3{}_3=-\frac{S^2}{ r^4}-\Lambda ,
\end{align}
where $S$ is a new constant introduced by the St\"{u}ckelberg field $\phi^a$.

\emph{SSS solution.} The field equation (\ref{ModEin}) reduce to one equation
\begin{equation}
  \frac{f'}{r}+\frac{f-1}{r^2}+m^2(-\frac{S^2}{ r^4}+\Lambda)+\frac{Q^2}{r^4}=0.
\end{equation}
The component $22$ of field equation (\ref{ModEin}) can be constraint by $\nabla_{\alpha}T^{(m)}{}^{\alpha\beta}=\nabla_{\alpha}T^{(\mathcal{K})}{}^{\alpha\beta}=0$.
It is easy to show that there is an solution
\begin{equation}
  f=1-\frac{2M}{r} +\frac{Q^2-m^2S^2}{r^2}-\frac{m^2\Lambda}{3}r^2.
\end{equation}
Parameter $M$ is the mass of black hole, the cosmological constant $\Lambda$ is related to the model parameters $\alpha_3, \alpha_4$. This solution is very similar to some solution given by Ref. \cite{Li2016-1}.

\emph{Newman-Janis Algorithm.} We now transform the SSS solution into a rotation case. Since there is a cosmological constant in the metric (\ref{StaticM}), we use the modified NJA given by Ref. \cite{Chaturvedi2023}. Firstly, we transform the metric (\ref{StaticM}) into the Eddington-Finkelstein coordinates by defining
\begin{equation}
  du=dt-\frac{dr}{f}.
\end{equation}
Thus, the metric (\ref{StaticM}) can be rewritten as
\begin{equation}\label{NullM}
 ds^2=fdu^2+2dudr-r^2(d\theta^2+\sin^2\theta d\varphi^2).
\end{equation}

Secondly, we choose a null tetrad $\{l^{\mu},n^{\mu},m^{\mu},\bar{m}^{\mu}\}$ based on the metric form (\ref{NullM}), where $\bar{m}^{\mu}$ means the complex conjugate of $m^{\mu}$. The null tetrad satisfy
\begin{align}
  l_{\mu}l^{\mu} &=m_{\mu}m^{\mu}=n_{\mu}n^{\mu}=l_{\mu}m^{\mu}=n_{\mu}m^{\mu}=0,  \\
  l_{\mu}n^{\mu} & =-m_{\mu}\bar{m}^{\mu}=1,
\end{align}
and
\begin{equation}
  g^{\mu\nu}=l^{\mu}n^{\nu}+l^{\nu}n^{\mu}-m^{\mu}\bar{m}^{\nu}-m^{\nu}\bar{m}^{\mu}.
\end{equation}
The null tetrad determined by metric (\ref{NullM}) have the well-known form 
\begin{align}
  l^{\mu} &=\delta^{\mu}_r  ,\\
  n^{\mu} &=\delta^{\mu}_u-\frac{f}{2}\delta^{\mu}_r,\\
  m^{\mu} &=\frac{1}{\sqrt{2}r}\left(\delta^{\mu}_{\theta}+\frac{i}{\sin\theta}\delta^{\mu}_{\varphi}\right).
\end{align}

Thirdly, we perform the complex transformation on the coordinates $\{u,r,\theta,\varphi\}$. As is shown in Ref. \cite{Azreg2014-1}, the metric functions $\{f(r),r^2\}$ are also transformed into $\{h(r,\theta,a),\Psi(r,\theta,a)\}$ at the same time. In the limit $a\rightarrow 0$, functions $\{h,\Psi\}$ are required to be reproduce to $\{f(r),r^2\}$, i.e.
\begin{align}
  \lim_{a\rightarrow 0}h(r,\theta,a) &=f(r),  \\
   \lim_{a\rightarrow 0}\Psi(r,\theta,a) & =r^2.
\end{align}
The usual NJA for the Einstein gravity involves the complex transformation of only $\{u,r\}$ coordinates. However, in the case of a cosmological constant, one requires an additional complexification of $\varphi$ coordinate as well. We consider the following transformation
\begin{align}
  du &\rightarrow du+ia\frac{\sin \theta}{\Delta_{\theta}}d\theta,  \\
  dr & \rightarrow dr-i a\sin\theta d\theta,\\
  d\varphi&\rightarrow d\varphi+\frac{ia}{\sin \theta}\left(-1+\frac{1+\frac{\Lambda m^2 a^2}{3}}{\Delta_{\theta}}\right)d\theta,
\end{align}
and $\theta$ did not do the transformation. In the case of Einstein gravity with a cosmological constant, the function $\Delta_{\theta}$ is given by
\begin{equation}
  \Delta_{\theta}=1+\frac{\Lambda m^2 a^2}{3}\cos^2\theta.
\end{equation}
Then, the $\delta^{\mu}_{\nu}$ transform as
\begin{align}
\delta^{\mu}_{u}&\rightarrow \delta^{\mu}_{u},\quad \delta^{\mu}_{r}\rightarrow \delta^{\mu}_{r},\nonumber\\
 \delta^{\mu}_{\theta}&\rightarrow ia\sin\theta(\frac{\delta^{\mu}_{u}}{\Delta_{\theta}}-\delta^{\mu}_{r})+\delta^{\mu}_{\theta}+\frac{i a}{\sin\theta}\left(-1+\frac{1+\frac{\Lambda m^2 a^2}{3}}{\Delta_{\theta}}\right)\delta^{\mu}_{\varphi},\quad \delta^{\mu}_{\varphi}\rightarrow\delta^{\mu}_{\varphi}.
\end{align}
Thus, the tetrad $(l^{\mu},n^{\mu},m^{\mu})$ are transformed into
\begin{align}
  l^{\mu} &=\delta^{\mu}_r  ,\label{null1}\\
  n^{\mu} &=\delta^{\mu}_u-\frac{h}{2}\delta^{\mu}_r,\\
  m^{\mu} &=\frac{1}{\sqrt{2\Psi}}\left(ia\sin\theta(\frac{\delta^{\mu}_{u}}{\Delta_{\theta}}-\delta^{\mu}_{r})+\delta^{\mu}_{\theta}+\frac{i a}{\sin\theta}\frac{1+\frac{\Lambda m^2 a^2}{3}}{\Delta_{\theta}}\delta^{\mu}_{\varphi}\right).\label{null2}
\end{align}
In the cosmological constant case \cite{Chaturvedi2023}, functions $h$ and $\Psi$ are defined by
\begin{align}
  h &=\frac{\Delta_r(r)-a^2\sin^2\theta \Delta_{\theta}(\theta)}{\rho^2}, \\
  \Psi & =\frac{\rho^2}{\Delta_{\theta}},
\end{align}
where $\rho^2=r^2+a^2\cos^2\theta$ and
\begin{equation}
  \Delta_{r}=(r^2+a^2)\left(1-\frac{\Lambda m^2 r^2}{3} \right)-2Mr+Q^2-m^2S^2.
\end{equation}

The forth step is to calculated the metric based on null tetrad (\ref{null1}) - (\ref{null2}). The direct calculations show
\begin{align}
  ds^2 &= h du^2+2dudr+\frac{2a\sin^2\theta}{1+\frac{\Lambda m^2 a^2}{3}}\left( \Delta_{\theta}-h \right)dud\varphi\nonumber \\
   & -\frac{2a\sin^2\theta}{1+\frac{\Lambda m^2 a^2}{3}}drd\varphi-\frac{\rho^2}{\Delta_{\theta}}d\theta^2-\frac{(a^2+r^2)^2\Delta_{\theta}-a^2\Delta_r\sin^2\theta}{\rho^2(1+\frac{\Lambda m^2 a^2}{3})^2}\sin^2\theta d\varphi^2.\label{NullM2}
\end{align}

The last step is to transform the metric (\ref{NullM2}) into the Boyer-Lindquist coordinate form. Define the following transformation
\begin{align}
   du & =dt-\frac{a^2+r^2}{\Delta_r}dr, \\
   d\varphi & =d\varphi-\frac{a}{\Delta_r}(1+\frac{\Lambda m^2 a^2}{3})dr,
\end{align}
we obtain the final results
\begin{align}
  ds^2 &=\frac{\Delta_r-a^2\sin^2\theta \Delta_{\theta}}{\rho^2}dt^2+\frac{2a\sin^2\theta((a^2+r^2)\Delta_{\theta}-\Delta_r)}{\rho^2(1+\frac{\Lambda m^2 a^2}{3})} dtd\varphi\nonumber \\
   & -\frac{\rho^2}{\Delta_r}dr^2-\frac{\rho^2}{\Delta_{\theta}}d\theta^2-\frac{(a^2+r^2)^2\Delta_{\theta}-a^2\Delta_r\sin^2\theta}{\rho^2(1+\frac{\Lambda m^2 a^2}{3})^2}\sin^2\theta d\varphi^2.\label{Kerr-Li}
\end{align}
One can also define $d\tilde{\varphi}=(1+\frac{\Lambda m^2 a^2}{3}) d\varphi$ to obtain a more concise form.

\emph{Conclusion.} We have derived the black hole family beyond Kerr-Newman in dRGT massive gravity. To ensure that the rotation metric (\ref{Kerr-Li}) is indeed the solution, we perform a rigorous mathematical derivation in another paper. From this solution (\ref{Kerr-Li}), we have observed for the first time that the graviton's mass term modifies the black hole's charge term in term $\frac{1}{r^2}$. The reason for this modification remains unclear. Additionally, we note that there are hairy solutions that only appear in cases of spherical symmetry \cite{Li2016-1}. It remains uncertain whether this is the only rotating solution available. The uniqueness of rotating black holes will be a focus of future research.

\textbf{Acknowledgement}
This work is partially supported by the National Natural Science Foundation of China (NSFC U2031112) and the Scientific Research Foundation of Hunan University of Arts and Sciences (23BSQD237). We also acknowledge the science research grants from the China Manned Space Project with NO. CMS-CSST-2021-A06.

\appendix
\section{The expressions of $T^{(\mathcal{K})}{}^{\alpha}{}_{\beta}$}
It can be calculated directly to show
\begin{align}
  \text{Det} \sqrt{\Xi_1} & =\sqrt{f_{11}f_{00}-f_{01}^2}, \\
  [\sqrt{\Xi_1}] &=\sqrt{\frac{f_{00}}{f}+f f_{11}+2(f_{00}f_{11}-f_{01}^2)},\\
  \text{Det} \sqrt{\Xi_2} & =\frac{\sqrt{f_{22}f_{33}}}{r^2\sin\theta},\\
  [\sqrt{\Xi_1}] &=\frac{\sqrt{f_{22}}}{r}+\frac{\sqrt{f_{33}}}{r\sin\theta}.
\end{align}
Define $c_3=3\alpha_3+12\alpha_4,c_4=1+6\alpha_3+12\alpha_4$, one obtain
\begin{align}
 T^{(\mathcal{K})}{}^{0}{}_{0} & =\frac{1}{2[\sqrt{\Xi_1}]}\bigg([\sqrt{\Xi_1}](-3 + 2 [\sqrt{\Xi_2}]- 2 c_3 ([\sqrt{\Xi_2}]-1)  + c_4 (-3 - 2 \text{det} ~\sqrt{\Xi_2} + 4 [\sqrt{\Xi_2}]) )\nonumber\\&+([\sqrt{\Xi_1}]^2+\sqrt{[\sqrt{\Xi_1}]^4-4[\sqrt{\Xi_1}]^2\text{det} ~\sqrt{\Xi_1}-4f_{01}^2 })(1+ c_3(\text{det} ~\sqrt{\Xi_2}-1)+c_4 ( 2-[\sqrt{\Xi_2}])) \bigg),\label{TK1} \\
 T^{(\mathcal{K})}{}^{1}{}_{1}& =\frac{1}{2[\sqrt{\Xi_1}]}\bigg([\sqrt{\Xi_1}](-3 + 2 [\sqrt{\Xi_2}]- 2 c_3 ([\sqrt{\Xi_2}]-1)  + c_4 (-3 - 2 \text{det} ~\sqrt{\Xi_2} + 4 [\sqrt{\Xi_2}]) )\nonumber\\&+([\sqrt{\Xi_1}]^2-\sqrt{[\sqrt{\Xi_1}]^4-4[\sqrt{\Xi_1}]^2\text{det} ~\sqrt{\Xi_1}-4f_{01}^2 })(1+ c_3(\text{det} ~\sqrt{\Xi_2}-1)+c_4 ( 2-[\sqrt{\Xi_2}]))\bigg), \\
 T^{(\mathcal{K})}{}^{2}{}_{2} & =\frac{1}{2[\sqrt{\Xi_2}]}\bigg([\sqrt{\Xi_2}](-3 + 2 [\sqrt{\Xi_1}]- 2 c_3 ([\sqrt{\Xi_1}]-1)  + c_4 (-3 - 2 \text{det} ~\sqrt{\Xi_1} + 4 [\sqrt{\Xi_1}]) )\nonumber\\&+([\sqrt{\Xi_2}]^2+\sqrt{[\sqrt{\Xi_2}]^4-4[\sqrt{\Xi_2}]^2\text{det} ~\sqrt{\Xi_2} })(1+ c_3(\text{det} ~\sqrt{\Xi_1}-1)+c_4 ( 2-[\sqrt{\Xi_1}])) \bigg),\\
 T^{(\mathcal{K})}{}^{3}{}_{3} & =\frac{1}{2[\sqrt{\Xi_2}]}\bigg([\sqrt{\Xi_2}](-3 + 2 [\sqrt{\Xi_1}]- 2 c_3 ([\sqrt{\Xi_1}]-1)  + c_4 (-3 - 2 \text{det} ~\sqrt{\Xi_1} + 4 [\sqrt{\Xi_1}]) )\nonumber\\&+([\sqrt{\Xi_2}]^2-\sqrt{[\sqrt{\Xi_2}]^4-4[\sqrt{\Xi_2}]^2\text{det} ~\sqrt{\Xi_2} })(1+ c_3(\text{det} ~\sqrt{\Xi_1}-1)+c_4 ( 2-[\sqrt{\Xi_1}])) \bigg),\\
 T^{(\mathcal{K})}{}^{0}{}_{1} & =\frac{-f_{01}}{[\sqrt{\Xi_1}]}(1+ c_3(\text{det} ~\sqrt{\Xi_2}-1)+c_4 ( 2-[\sqrt{\Xi_2}])).
\end{align}
Solving $ T^{(\mathcal{K})}{}^0{}_0=T^{(\mathcal{K})}{}^{1}{}_{1}=\frac{S^2}{ r^4}-\Lambda ,T^{(\mathcal{K})}{}^2{}_2=T^{(\mathcal{K})}{}^3{}_3=-\frac{S^2}{ r^4}-\Lambda,T^{(\mathcal{K})}{}^0{}_1=0$, one obtain $f_{00},f_{01},f_{11},f_{22},f_{33}$. Five equations will uniquely determine a set of five unknowns.


\end{document}